\documentclass[iop,apjl]{emulateapj}
\usepackage{apjfonts}

\renewcommand{\baselinestretch}{1.05}

\begin{document}

\title{Evidence for Enrichment by Supernovae in the Globular Cluster NGC 6273}
\shorttitle{NGC 6273}
\shortauthors{Han et al.}

\author{Sang-Il Han\altaffilmark{1,2,3},
 Dongwook Lim\altaffilmark{1,3},
 Hyunju Seo\altaffilmark{1},
 Young-Wook Lee\altaffilmark{1,4}
}

\altaffiltext{1}{Center for Galaxy Evolution Research and Department of Astronomy, Yonsei University, Seoul 03722, Korea}
\altaffiltext{2}{Korea Astronomy and Space Science Institute, Daejeon 34055, Korea}
\altaffiltext{3}{Both authors have contributed equally to this paper.}
\altaffiltext{4}{Corresponding author: ywlee2@yonsei.ac.kr}

\begin{abstract}
In our recent investigation \citep{lim15}, we have shown that narrow-band photometry can be combined with low-resolution spectroscopy to  effectively search for globular clusters (GCs) with supernovae (SNe) enrichments. Here we apply this technique to the metal-poor bulge GC NGC~6273, and find that the red giant branch stars in this GC are clearly divided into two distinct subpopulations having different calcium abundances. The Ca rich subpopulation in this GC is also enhanced in CN and CH, showing a positive correlation between them. This trend is identical to the result we found in M22, suggesting that this might be a ubiquitous nature of GCs more strongly affected by SNe in their chemical evolution. Our results suggest that NGC~6273 was massive enough to retain SNe ejecta which would place this cluster in the growing group of GCs with Galactic building block characteristics, such as $\omega$~Centauri and Terzan~5.
\end{abstract}


\section{Introduction}
One of the problems of the $\Lambda$CDM hierarchical merging paradigm in local scale is that the observed number of the remaining ``building blocks'' in the Milky Way is about an order of magnitude smaller than the predicted number of subhalos \citep[see, e.g.,][]{kau93,moo06}. Despite the recent progress in searching for the ultra faint dwarf galaxies \citep[e.g.,][references therein]{wil05,zuc06,bel07,bec15}, this ``missing satellite problem'' still requires additional search for the building block candidates in the Milky Way. Not well known to the community is a possibility that some massive globular clusters (GCs) might be originated from proto-galactic subsystems \citep{sea78,fre93}. Starting from  $\omega$~Centauri \citep{lee99,bed04}, increasing number of GCs now show evidence of supernovae (SNe) enrichment, indicating that their original mass was heavy enough to retain SNe ejecta \citep{dop86,bau08}, which would place them as remaining nuclei of disrupted dwarf galaxies. As of August 2015, GCs belong to this category include $\omega$~Cen \citep{lee99,bed04}, M54 \citep{sie07}, M22 \citep{jlee09,dac09,mar09}, Terzan~5 \citep{fer09}, NGC~1851 \citep{jlee09,han09,car10}, NGC~2419 \citep{coh12,muc12,lee13}, NGC~5824 \citep{dac14}, M2 \citep{yon14}, and NGC~5286 \citep{mar15}.  Although it is widely accepted by
the community that almost all GCs were initially much more massive \citep[e.g.,][]{dec07,der08,car10a,con11,sch11}, these GCs that were able to retain SNe ejecta were probably even more massive \citep[M $>$ 10$^{7.5}$M$_{\odot}$;][]{dec10}, and therefore, could have provided more stars to halo field.

More massive GCs in the Milky Way bulge are expected to have this characteristic \citep{die07,lee07,fer09,lim15}. Detailed studies for these GCs, however, are hampered by highly obscured environment in which they are placed. The near IR photometry and spectroscopy would be ideal probes to detect any evidence for multiple populations in these GCs as shown, for example, by \citet{fer09} for the case of Terzan~5. As suggested by \citet[][hereafter Paper I]{lim15}, in the case of mildly obscured GCs in the bulge, however, the combination of narrow-band Ca photometry and low-resolution spectroscopy can effectively disentangle the reddening and metallicity effects. The purpose of this letter is to report our discovery, by employing this technique, that NGC~6273 also belongs to this growing body of GCs. \\

\begin{figure*} 
\includegraphics[scale=0.8]{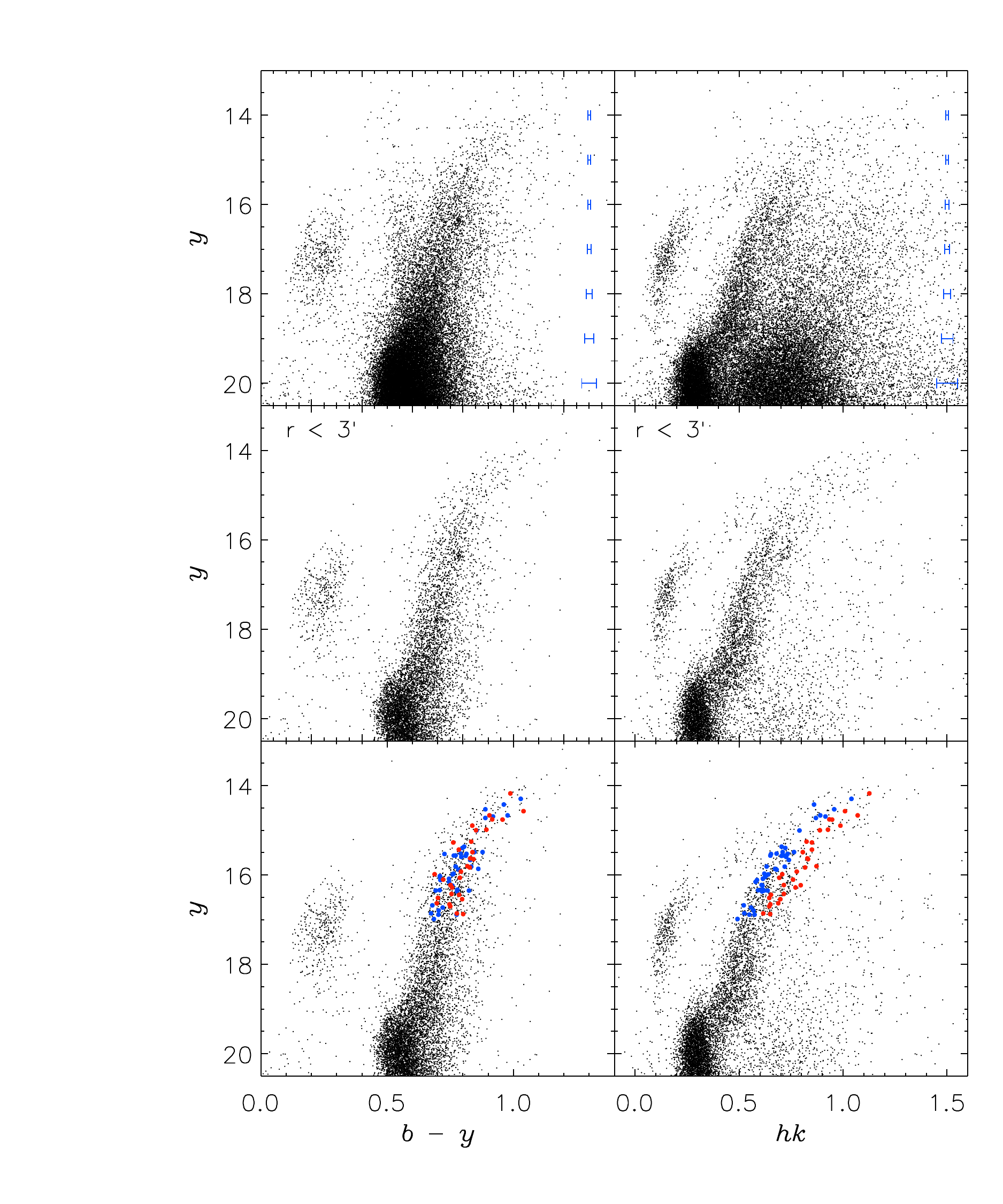}
\caption{CMDs for NGC~6273 in  ($y, b-y$) and ($y, hk$) planes. The top panels show all stars while the middle panels are for stars within 3$\arcmin$. The horizontal bars denote the measurement error ($\pm$1$\sigma$). Spectroscopic targets are identified on the bottom panels, where blue and red circles denote stars selected from the bluer and redder RGBs, respectively.}

\label{f:cmd}
\end{figure*}

\begin{figure*} 
\includegraphics[scale=0.7,angle=90]{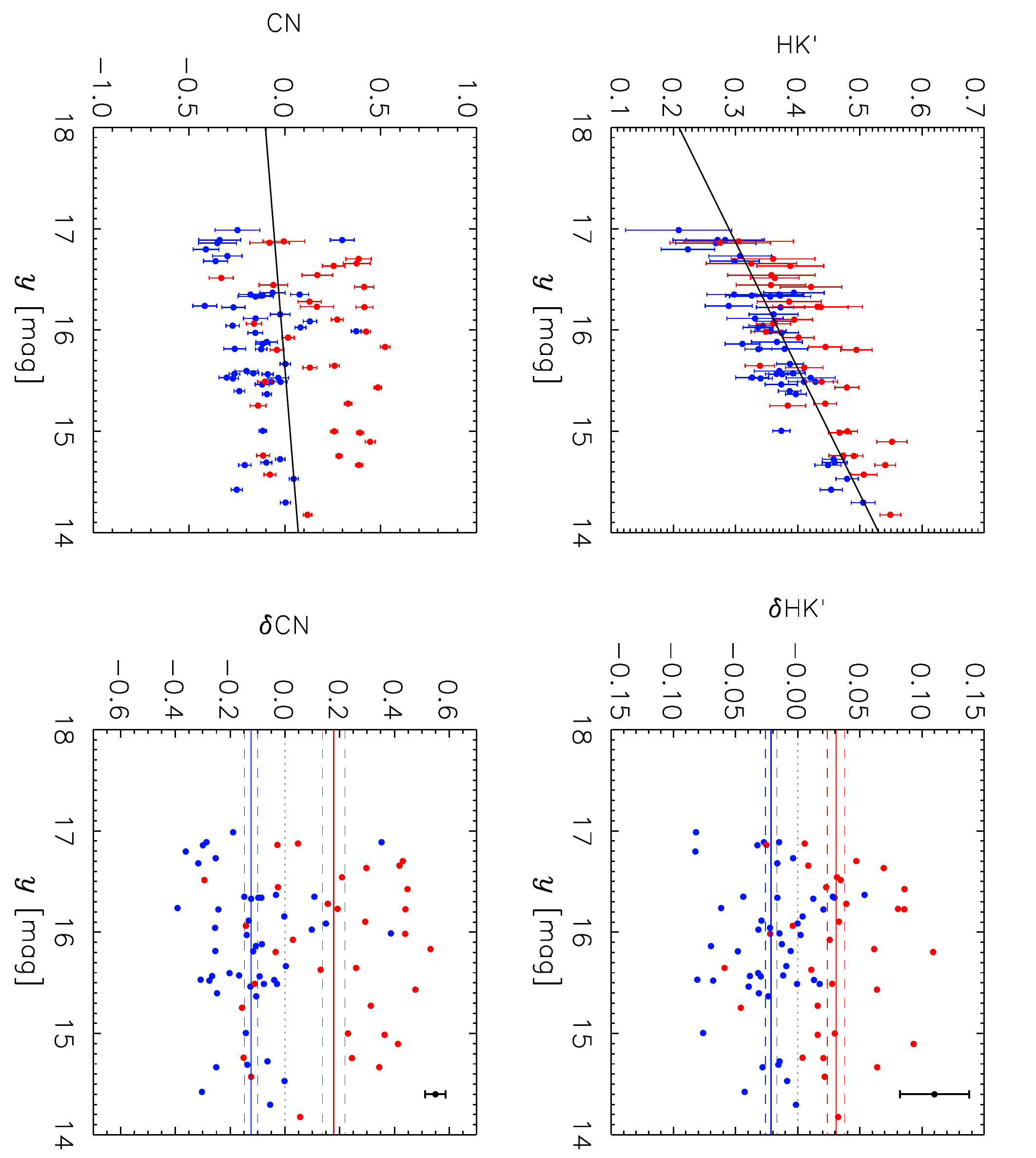}

\caption{Measured HK$^\prime$ (upper panels) and CN (lower panels) indices as functions of $y$ magnitude. The $\delta$HK$^\prime$ and $\delta$CN indices in the right panels are obtained from the height of the HK$^\prime$ and CN indices above the least-square lines in the left panels. The blue and red circles are stars in bluer and redder RGBs in the bottom panels of Figure~\ref{f:cmd}. The solid and dashed lines indicate the mean value and the standard deviation of the mean ($\pm$1$\sigma$) for each subpopulation, respectively. The vertical bars in the right panels show the typical measurement error for $\delta$HK$^\prime$ and $\delta$CN indices.
(Supplemental data for this figure are available in the online journal.) }

\label{f:spc} 
\end{figure*}

\section{Presence of two red giant branches with different Ca abundances}
Our observations, both narrow-band Ca photometry and low-resolution spectroscopy, were made on the du Pont 2.5m telescope at Las Campanas Observatory during the 14 nights from 2011 to 2014. The narrow-band Ca photometry covered 15{\arcmin} $\times$ 15{\arcmin} field centered on the cluster. Total exposure times for the central region (9{\arcmin} $\times$ 9{\arcmin}) were 12600, 2520, and 1260 sec for Ca, $b$, and $y$ filters, respectively, and those for the outer regions were 3600, 720, and 360 sec. 
The low-resolution spectroscopy was performed on the multi-object spectroscopic mode of the Wide Field Reimaging CCD (WFCCD) in the same field covered by the photometry. The exposure times were between 1200 and 1800 sec, depending on the magnitude of the target stars in the masks. For the data reduction, both for the photometry and spectroscopy, we followed the same manner described in Paper I -- IRAF\footnote{IRAF is distributed by the National Optical Astronomy Observatory, which is operated by the Association of Universities for Research in Astronomy (AURA) under a cooperative agreement with the National Science Foundation.} for the preprocessing, DAOPHOT II/ALLSTAR and ALLFRAME \citep{ste87,ste94} for the point spread function photometry, and the modified WFCCD reduction package (\citealt{pro06}; Paper I) for the spectroscopic analysis. Readers are referred to Paper I for the details of procedures adopted in our data reduction. 

Figure~\ref{f:cmd} shows color-magnitude diagrams (CMDs) for NGC~6273 in ($y, b-y$) and ($y, hk$) planes, where the $hk$ index is defined as $hk = ($Ca$-b)-(b-y)$ \citep{att91}. As described in Paper I, stars affected by adjacent starlight contamination and bad measurements were rejected (i.e., $sep$ $< $0.5, $\chi$ $>$ 1.5, and SHARP $>$ $|0.25|$). We adopted more rigorous criterion as to the $sep$ index (less than 0.5 instead of 1.0 as adopted in Paper I) because of a larger field star contamination in the bulge field where NGC 6273 is placed. The top panels of Figure~\ref{f:cmd} show all stars in our program field, where the field star contamination (mostly main-sequence stars in the disk and red giant branch stars in the bulge) is evident. The stars within 3{\arcmin} from the cluster center are shown in the middle panels where the CMD features from the cluster members are more clearly visible.  In these panels, the most notable feature is the presence of two distinct red giant branch (RGB) subpopulations in the ($y, hk$) CMD, while this split is not apparent in the ($y, b-y$) CMD, suggesting a difference in calcium abundance between the two RGB sequences. Note that this spilt is also present on the sub giant branch in the ($y, hk$) CMD. The difference in the $hk$ index between the two RGBs is about 0.12 mag at the magnitude level of the horizontal-branch. While some field star contamination is still present\footnote{Photometry for a control field obtained in the same observing run which is two tidal radius away from the cluster center shows field star contamination is small (less than 6\%) in the regime occupied by two RGBs in the mag range of 16 $< y <$ 17.}, the population ratio between the bluer and redder RGBs, defined by the difference in $\Delta$$hk$ from the fiducial line ($\Delta$$hk$ $<$ 0.06 for bluer RGB, $\Delta$$hk$ $\geq$ 0.06 for redder RGB), is estimated to be 0.6:0.4 for the stars in the mag range of 16 $< y <$ 17.

In order to confirm the difference in calcium abundance between the two RGBs, we have measured  spectroscopic HK$^\prime$, $\delta$HK$^\prime$, CN, and $\delta$CN indices defined in Paper I for a similar number of stars selected from each RGB. The target stars are shown in the bottom panels of Figure~\ref{f:cmd}, and Figure~\ref{f:spc} shows measured indices as functions of $y$ magnitude. The right panels of Figure~\ref{f:spc} are the differences in HK$^\prime$ and CN indices ($\delta$HK$^\prime$ and $\delta$CN) measured from the least-square fits (black solid lines in left panels) to show only the abundance effect at fixed temperature and gravity (see Paper I). The mean differences in $\delta$HK$^\prime$ and $\delta$CN indices between the two subpopulations are 0.05 dex and 0.31 dex, respectively, in the sense that the redder RGB stars are more enhanced. These differences are significant at 6.3$\sigma$ level for $\delta$HK$^\prime$  and at 6.4$\sigma$ level for $\delta$CN compared to the standard deviation of the mean for each group. 

\begin{figure} 
\includegraphics[scale=0.55,angle=90]{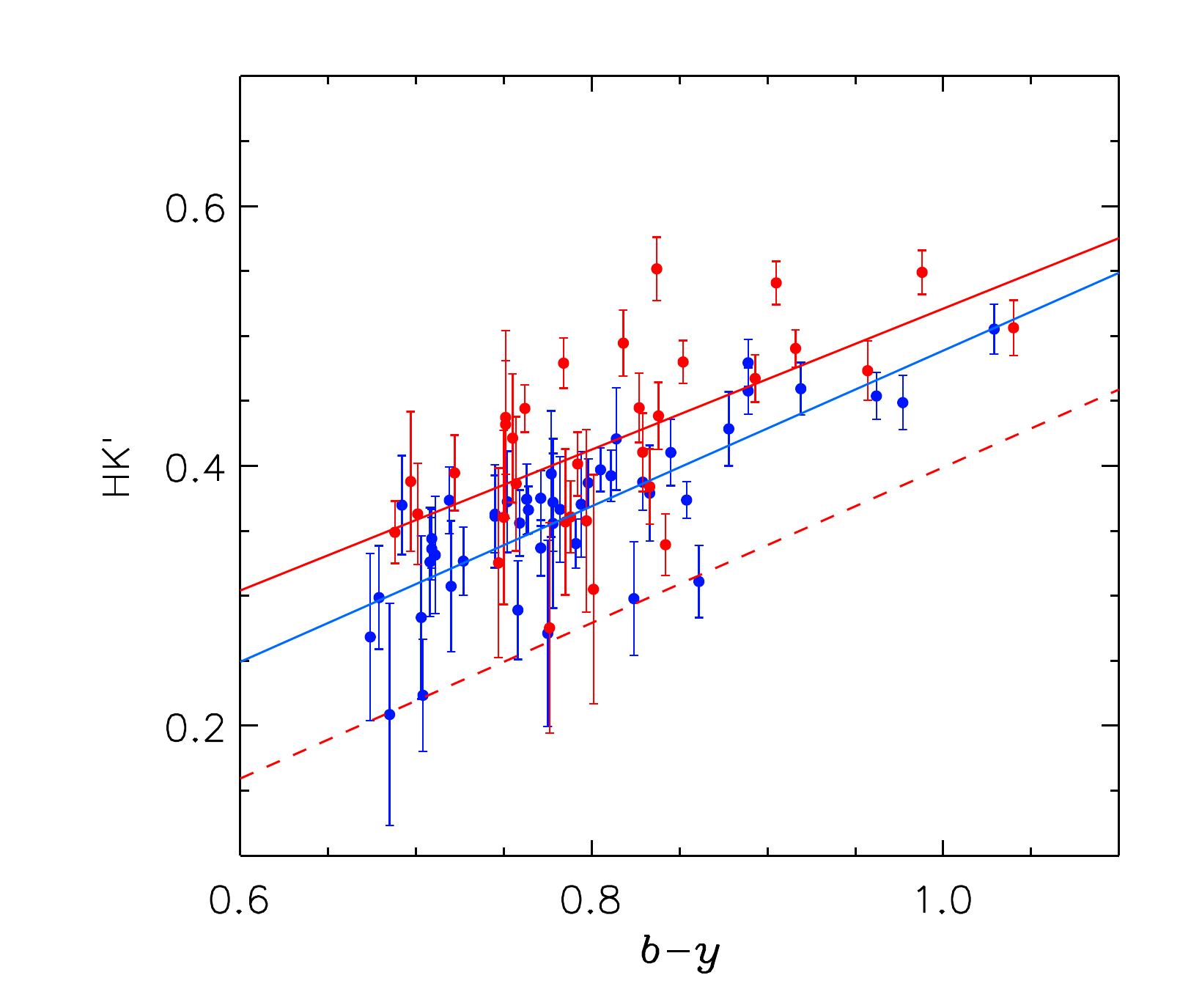}

\caption{Measured HK$^\prime$ index as a function of $b-y$ color. The blue and red solid lines are the least-square lines for each subpopulation, respectively. The red dashed line indicates the expected locus of redder RGB stars if the RGB split is due to the reddening.}
\label{f:redding}
\end{figure}

Since some differential reddening  is reported in the field toward NGC~6273 ($\Delta$$E(B-V )$$\sim$0.2; \citealt{pio99}), it is important to check whether the RGB split is affected by the reddening. In order to examine this, we plot HK$^\prime$ index versus $b-y$ color in Figure~\ref{f:redding}. If the RGB split is due to the reddening, the redder RGB stars would be placed on the right side in this diagram or around the red dashed line. However, the observed redder RGB stars are located in the opposite regime as would be expected if they are enhanced in HK$^\prime$. This confirms, together with the discrete distribution of the two RGB sequences on the CMD, that the effect from the differential reddening, if any, is not significant. Furthermore, note that the $hk$ index has a little dependence on the reddening, $E(hk)/E(B-V)=-0.12$ and $E(hk)/E(b-y)=-0.16$ \citep{att91,jlee09}. \\

\begin{figure}
\centering
\includegraphics[scale=0.6]{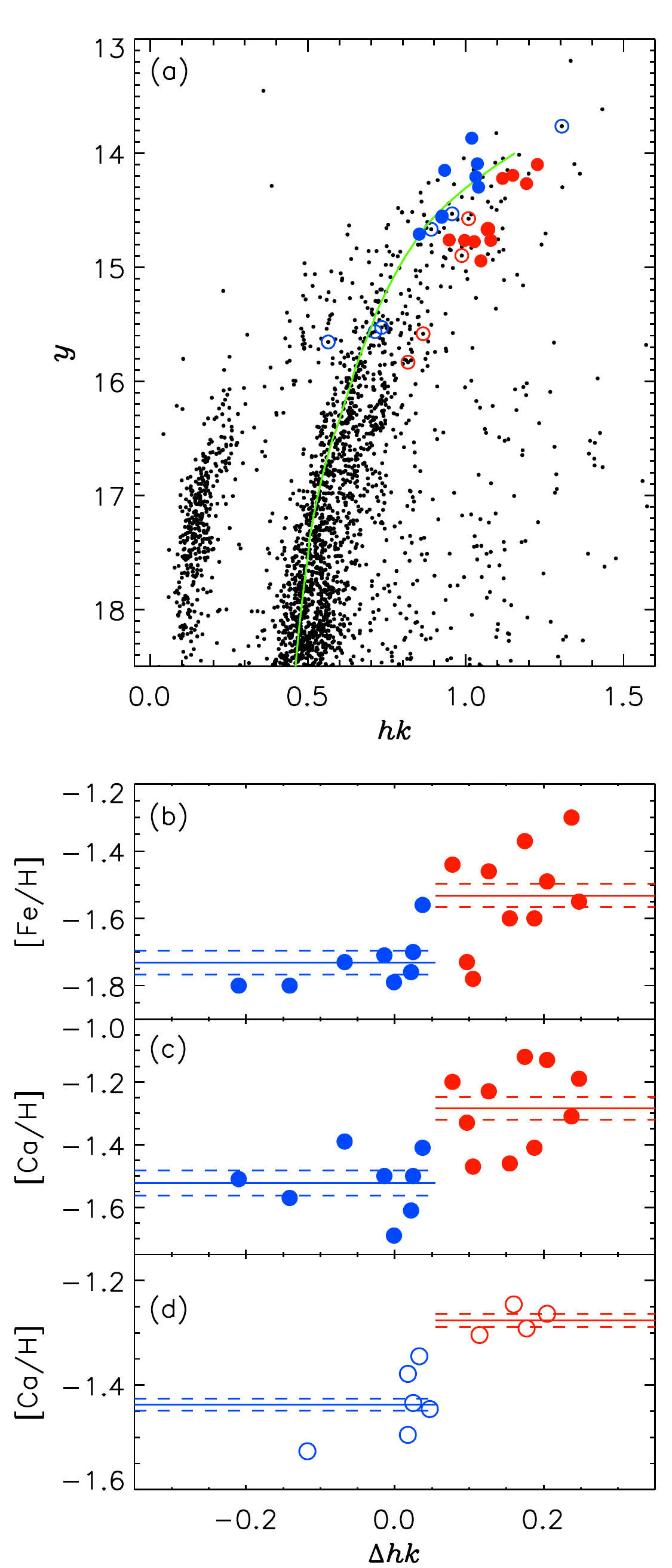}
\caption{Difference in metallicity between the two RGB subpopulations in NGC~6273. Blue and red circles denote bluer and redder RGB stars, respectively, for which cross-matched spectroscopic data are available in \citet[][filled circles]{joh15} and \citet[][open circles]{rut97}. The redder RGB stars are enhanced in [Fe/H] and [Ca/H] (see panels (b) and (c)). The $\Delta$$hk$ index in panels (b)-(d) is defined as the difference from the fiducial line (green solid line in panel (a)). The solid and dashed lines indicate, respectively, the mean value and $\pm$1$\sigma$ error of the mean for each subpopulation.}

\label{f:comspc}
\end{figure}

\section{Discussion}
We have shown that the RGB stars in NGC~6273 are divided into two distinct subgroups in the narrow-band Ca photometry. Our low-resolution spectroscopy confirms that the origin of this split is indeed due to the difference in calcium abundance between the two RGBs.  After the draft of this paper was completed, the iron spread in this GC was recently reported by \citet{joh15} from high-resolution spectroscopy of 18 RGB stars. While this result is qualitatively consistent with our finding, the discrete nature of the metallicity distribution function is more clearly shown in this investigation from the narrow-band Ca photometry for $\sim$1,600 RGB stars, together with the low-resolution spectroscopy for 78 stars.

In order to see whether the calcium enhanced stars in our sample are similarly enhanced in iron abundance, in panel (a) of Figure~\ref{f:comspc}, we have cross-matched stars in \citet{joh15} with those in our sample (filled circles). The two subpopulations were defined from the photometry by the difference in $\Delta$$hk$ from the fiducial line as described in Section 2. Panel (b) shows that the calcium enhanced stars in our observations are enhanced in [Fe/H] as well, with the difference of 0.2 dex (4.0$\sigma$ level). Calcium abundance obtained from Johnson et al. (2015) in panel (c) also confirms the difference in [Ca/H] (0.24 dex, 4.4$\sigma$ level).\footnote{The two subpopulations also appear to show the difference in [Mg/Fe]. However, when [Mg/H] is plotted, instead of [Mg/Fe], we found no difference between the two groups as correctly pointed out by the referee. This confirms that the apparent difference in [Mg/Fe] is only the reflex of the variation in Fe between the two subpopulations.} We also compare the Ca II triplet  measurements of \citet{rut97} with our photometry (open circles), where the difference in [Ca/H] was deduced from their reduced equivalent widths ($\Sigma$Ca$_{\rm HB}$). In this case, the mean difference in [Ca/H] between the two subpopulations is 0.16 dex, which is significant at 9.3$\sigma$ level (see panel (d)). 

\begin{figure} 
\includegraphics[scale=0.35,angle=90]{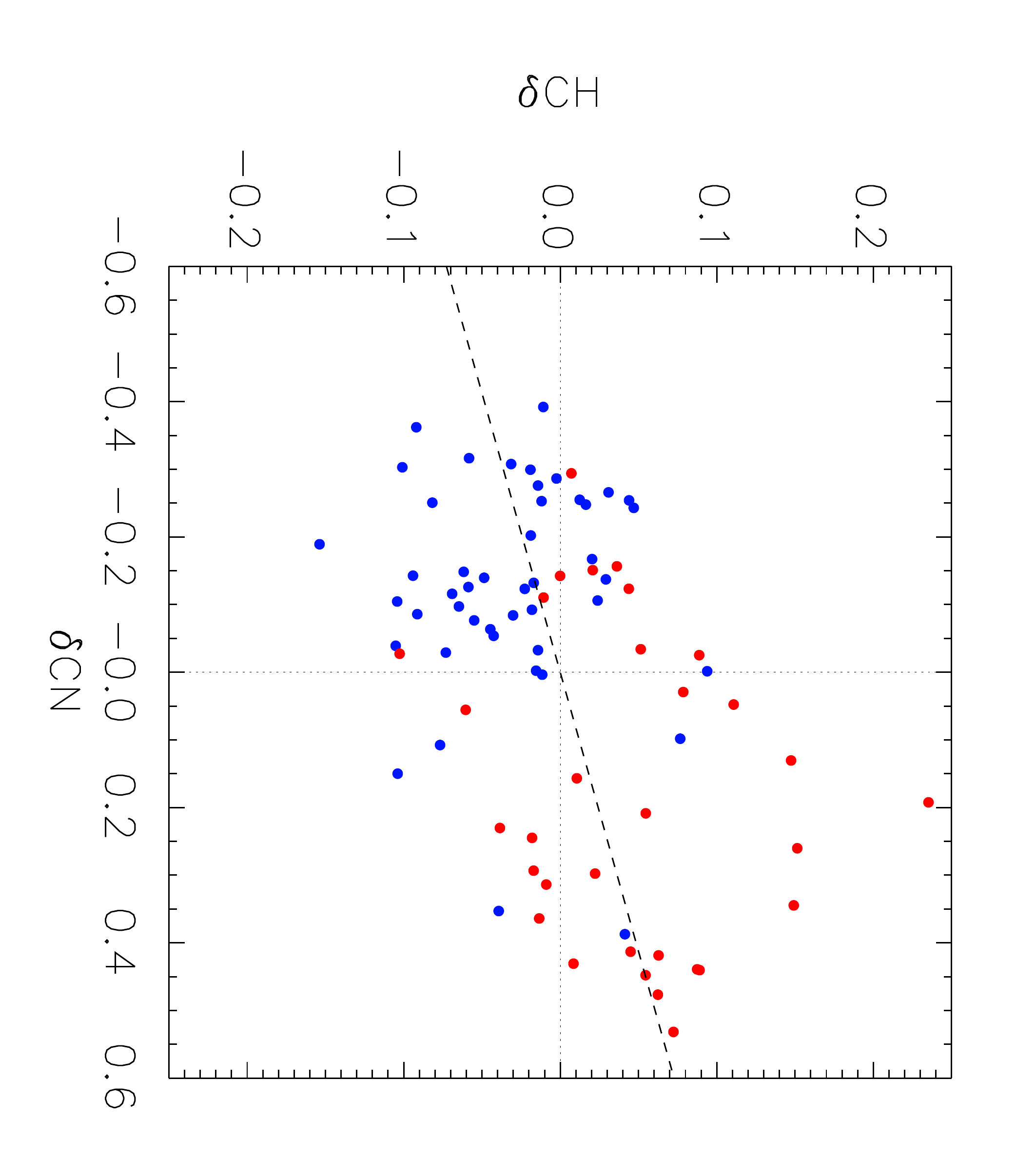}

\caption{The correlation between CN and CH indices. The RGB stars in NGC~6273 show a positive correlation, which is qualitatively similar to the case of M22. The dashed line is a least-square fit.}

\label{f:cnch}
\end{figure}

Figure~\ref{f:cnch} presents the correlation between $\delta$CN and $\delta$CH for the RGB stars in NGC~6273, which shows a positive correlation.  As discussed in Paper I (see Figure 13), most ``normal'' GCs show anti-correlation between CN and CH, while some massive GCs enriched by SNe, such as M22, show positive correlation. It was suggested in Paper I that this systematic difference would reflect how strongly SNe have enriched the later generation of stars. The fact that NGC~6273 is following the same trend observed in M22 is also consistent with our conclusion that the chemical evolution in this GC was relatively strongly affected by SNe. This further suggests that NGC~6273 was massive enough to retain SNe ejecta, which would place this cluster in the growing group of GCs with Galactic building block origin, such as $\omega$~Cen and Terzan~5. In particular, this places NGC~6273 as the first metal-poor ([Fe/H] $<$ -1.5) GC in the bulge field with this characteristic.

\acknowledgments{We are indebted to the anonymous referee for a number of helpful suggestions which led to several improvements in the manuscript. We also thank the staff of LCO for their support during the observations. Support for this work was provided by the National Research Foundation of Korea to the Center for Galaxy Evolution Research and by the Korea Astronomy and Space Science Institute under the R\&D program (No. 2014-1-600-05) supervised by the Ministry of Science, ICT and future Planning.}

\end{document}